\documentstyle[preprint,aps]{revtex}
\tighten
\draft
\widetext
\input epsf
\preprint{NYU-TH-99/07/03, HUTP-99/A040, NUB 3204}
\bigskip
\bigskip
\begin{document}
\title{A Remark on Witten  Effect for QCD Monopoles in Matrix Quantum 
Mechanics}
\medskip

\author{Gregory Gabadadze$^1$\footnote{E-mail: gabadadze@physics.nyu.edu} 
and Zurab Kakushadze$^{2,3}$\footnote{E-mail: 
zurab@string.harvard.edu. Address after September 1, 1999: C.N. Yang
Institute for
Theoretical Physics, State University of New York, Stony Brook, NY 11794.}}

\bigskip
\address{
$^1$Department of Physics, New York University, New York, NY 10003\\
$^2$Jefferson Laboratory of Physics, Harvard University,
Cambridge,  MA 02138\\
$^3$Department of Physics, Northeastern University, Boston, MA 02115}

\date{August 4, 1999}
\bigskip
\medskip
\maketitle

\begin{abstract}
In a recent work (hep-th/9905198) 
we argued that  a certain matrix quantum 
mechanics may describe  't Hooft's monopoles  which
emerge in  QCD when the theory is projected 
to its maximal Abelian subgroup.
In this note we find further  evidence which supports this 
interpretation. 
We study the theory  with a non-zero theta-term. In this case, 
't Hooft's QCD monopoles become dyons  since 
they acquire electric charges due to the Witten effect. 
We calculate a potential between a dyon  and
an anti-dyon  in the matrix quantum mechanics, and  
find that the attractive force between them grows  
as the theta angle increases.

\end{abstract}

't Hooft has shown \cite {tHooft} that a new and rich structure 
emerges in QCD  when certain unusual gauge fixing conditions
are imposed on the theory. This class of unitary gauges
projects QCD onto its maximal Abelian subgroup, {\em i.e.},
breaks $SU(N)$ down to $U(1)^{N-1}$. In addition to massless
and massive gauge fields, which are present after this symmetry breaking,
some point-like monopoles emerge in the theory \cite {tHooft}. 
These monopoles appear in a somewhat unusual way, 
as singularities of a chosen gauge fixing condition. 
Given the importance of these monopoles in various analytic and lattice 
studies of QCD (see \cite {rev} and references therein), 
it is desirable to have some gauge independent
description of  the dynamics of these objects. 
Recently, we addressed this issue in \cite {GK}
by considering pure QCD on a spatial three-torus. The 
T-dual form of pure QCD on a spatial torus can be 
interpreted as a certain matrix quantum mechanics. In \cite {GK} 
some arguments were presented which lead us to 
conjecture  that this quantum mechanics describes the  dynamics 
of 't Hooft's monopoles. 
The aim of this note is to seek for
further evidence in favor  of this conjecture.
Below we consider pure QCD with the theta term. It is known that
't Hooft's monopoles acquire electric charges due to the Witten effect
\cite {witten} once non-zero theta angle is 
introduced \cite {tHooft}. 
If the identification of the excitations of 
the matrix quantum mechanics with 't Hooft's monopoles is 
correct, then the Witten effect should also be seen 
within the matrix model. In other words, 
if one calculates the interaction force between the point-like objects
of the matrix model, then 
this force should depend on the theta angle. 
In fact, 
the attractive force between 
a  monopole and an antimonopole should  be greater then 
it is for a zero theta angle. 
In what follows we will show that this is indeed the case: 
An interacting monopole-antimonopole pair becomes a 
dyon-anti-dyon pair
once the theta angle is switched on, and the attractive 
force between them grows as $\theta$ increases. 
  
{}Consider four-dimensional pure $SU(N)$ QCD in the presence of the
$\theta$ angle. The corresponding Lagrangian density reads:
\begin{equation}\label{YM}
 {\cal L}_{\rm YM}=-{1\over 4 g^2_{\rm YM}} G_{\mu\nu}^a
 G^{a\mu\nu} +{\theta\over 32\pi^2} G_{\mu\nu}^a{\widetilde G}^{a\mu\nu}=
 -{1\over 4} {\rm Re}\left( \tau\left[G_{\mu\nu}^a
 G^{a\mu\nu} +i G_{\mu\nu}^a{\widetilde G}^{a\mu\nu}\right]\right)~.
\end{equation}
Here ${\widetilde G}^{a\mu\nu}\equiv{1\over 2}\epsilon^{\mu\nu\lambda\rho}
G_{\lambda\rho}^a$, and $\tau=\tau_1+i\tau_2\equiv 1/g_{\rm YM}^2
+i\theta/8\pi^2$. To avoid complications with Gribov copies, in the 
following we will be working in the  $A_0=0$ gauge.

{}Let $\zeta\equiv 1/\Lambda_{\rm YM}$ be the effective correlation length of
the theory, where $\Lambda_{\rm YM}$ is the dynamically generated QCD
scale. Let us compactify the theory on a rectangular three-torus $T_L\equiv
S^1\times S^1\times S^1$ with the radii of all three circles equal $L\gg
\zeta$. This corresponds to the strong coupling regime of the theory
(see discussions in \cite {GK}).
We can rewrite pure QCD compactified on $T_L$ as a
matrix quantum mechanics compactified on a dual three-torus $T_R\equiv
{\widetilde S}^1\times {\widetilde S}^1\times {\widetilde S}^1$ with the radii
of all three circles equal $R\equiv \alpha^\prime/L$, where the parameter
$\alpha^\prime$ is defined via the QCD scale $\Lambda_{\rm YM}$ as follows:
$\alpha^\prime\equiv \zeta^2=1/\Lambda_{\rm YM}^2$. In this T-dual
formulation of the theory the dynamical variables are time-dependent
matrices $\Phi_i(t)$, $i=1,2,3$, transforming in the adjoint representation
of (global) $SU(N)$. In addition to the color indices, for a given value of the
index $i=1,2,3$, the matrices $\Phi_i$ also carry indices corresponding to
the winding modes. In general,  
$\Phi$'s give a matrix representation of a covariant derivative on a torus
\cite {Taylor}. In the following we will suppress for simplicity
these winding indices. 
The corresponding Lagrangian of the matrix quantum mechanics 
(with the appropriate normalization for $\Phi_i$)
is given by:
\begin{equation}\label{lagrange}
{\cal L} =  {1\over 2  g \sqrt{\alpha'}}~ {\rm Tr}~ \left(
{\dot \Phi}^2_i +{1\over 2 (2\pi\alpha')^2}
\left[ \Phi_i,\Phi_j \right]^2 - {i \lambda\over 2\pi\alpha^\prime} 
 \epsilon_{ijk} \left[\Phi_i,\Phi_j\right]{\dot \Phi_k}
 \right)~,
\end{equation}  
where ${\dot \Phi}_i$ denotes the time derivative of $\Phi_i$, and the
traces over color and winding indices (with appropriate normalizations) 
are implicit. This Lagrangian should be amended by a corresponding
constraint equation (a counterpart of the Gauss's law) to describe
pure QCD in a T-dual picture \cite {GK}. 
The new coupling constant $g$ is
defined as follows:
\begin{equation}
g\equiv ({R/L})^{3/2} {g^2_{\rm YM}/4\pi}~.
\end{equation}
Also, the last term in (\ref{lagrange}) is due to the $\theta$-term in
(\ref{YM}), and the corresponding coupling $\lambda$ is given by:  
\begin{equation}
 \lambda\equiv\tau_2/\tau_1=\theta g_{\rm YM}^2/8\pi^2~.
\end{equation}
For $\theta=0$ (\ref{lagrange}) reduces to the usual bosonic matrix quantum
mechanics Lagrangian \cite{Witt,Danielson,KP,DKPS,BFSS,Taylor,Marty}.

{}Following \cite{GK} 
the Lagrangian (\ref{lagrange})
describes the dynamics of 't Hooft's QCD monopoles. 
Let us notice that in the limit $L\gg\zeta$, that is, 
$R\ll\zeta$, which we are interested in, 
the matrix quantum mechanics (\ref {lagrange})
is as  complicated a theory  as strongly coupled pure QCD, 
the reason being that it contains 
light {\em winding} modes (which map to the Kaluza-Klein modes in
the T-dual QCD  description) whose masses scale as $R/\alpha^\prime=1/L$
\cite {GK}.  
However, certain aspects of pure QCD in a large volume (which is a
strongly coupled theory) might be more transparent in the 
matrix quantum mechanics approach. In particular, 
the monopole mass in the theory is given by $M=1/2g\sqrt{\alpha'}$,
and in the regime we are discussing  $M\gg\Lambda_{\rm YM}$ \cite {GK}. 
Thus, it is reasonable to  consider 
interactions between monopoles when they  are moving very slowly
(or, are almost at rest).  At such a low  energies the light winding modes
are not yet excited in the model. Thus, we can 
neglect the contributions of these modes 
in the calculation.  
On the matrix model side interactions between monopoles 
(a monopole-antimonopole pair) are 
described by off-diagonal elements in $\Phi_i$. Thus,
following \cite {GK},  consider the $U(2)$
case where we have two monopoles with opposite magnetic charges. Let us make
the standard decomposition of the $\Phi$ field into it's classical and
quantum parts:
\begin{equation}
 \Phi_i=\Phi_i^{\rm cl}+\delta\Phi_i~.
\end{equation}
Here we choose the classical solution as follows:
\begin{eqnarray}
\Phi_1^{\rm cl}~=~{1\over 2}~\left (
\begin{tabular} {c c}
$r$ & 0 \\
0 &  $-r$ 
\end{tabular}
\right )~,~~~\Phi^{\rm cl}_2=0~,~~~\Phi^{\rm cl}_3=0~.
\end{eqnarray}
That is, the two monopoles are at a distance $r$ apart from each other 
in one of the spatial directions. 
In order to find an effective potential between them
one performs
integration with respect to off-diagonal fluctuations. 
In the absence of the $\theta$-angle the effective potential between the 
monopoles is given by (see \cite{GK} and references therein):
\begin{equation}\label{eff}
 V^{(\theta=0)}_{\rm eff}\propto {r\over \alpha^\prime}~.
\end{equation}
In fact, the easiest way to deduce the effective potential in 
the presence of the
$\theta$-angle is to rewrite  the Lagrangian (\ref{lagrange})
as follows: 
\begin{equation}
{\cal L} =  {1\over 2  g \sqrt{\alpha'}} ~{\rm Tr}~\left(
\left({\dot \Phi}_i -{i \lambda\over 2(2\pi\alpha^\prime)} 
 \epsilon_{ijk} \left[\Phi_i,\Phi_j\right]\right)^2
 +{1+\lambda^2\over 2 (2\pi\alpha')^2}
\left[ \Phi_i,\Phi_j \right]^2 \right)~.
\label{quadratic}
\end{equation}
Note that the difference between the Lagrangians with $\lambda=0$ and
$\lambda\not=0$ is in the redefinition of the conjugate momentum 
and rescaling
$\alpha^\prime\rightarrow\alpha^\prime/\sqrt{1+\lambda^2}$ in the
term containing the commutator $\left[\Phi_i,\Phi_j\right]^2$
(which is responsible for interactions between monopoles)\footnote{
This is, however, not true for the Hamiltonian of the theory
which will contain a term linear in theta along with the quadratic term  
arising  in front of the commutator $\left[\Phi_i,\Phi_j\right]^2$.}. 
Performing explicitly integration of the off-diagonal excitations
in (\ref {quadratic}), and calculating the corresponding functional
determinant,  
one finds 
the effective potential in the presence of the $\theta$-angle
\begin{equation}\label{eff1}
 V^{(\theta \ne 0)}_{\rm eff}~=~\sqrt{1+\lambda^2}~ 
V^{(\theta=0)}_{\rm eff}~.
\end{equation}
We see that, as in the case without the $\theta$-angle, there is a
linearly rising potential between a monopole and an anti-monopole. Thus,
there is a string stretched between them, and the string tension $T_s$ 
has a non-trivial $\theta$-dependence:
\begin{equation}\label{tension}
 T_s\propto \sqrt{1+\lambda^2}~\Lambda_{\rm YM}^2=
 \sqrt{1+\left(\theta g_{\rm YM}^2/8\pi^2\right)^2}~\Lambda_{\rm YM}^2~.
\end{equation}
As a result, the attraction force between 
the pair increases when the theta angle is switched on.
This corresponds to the fact that monopoles acquire
electric charges and become dyons. 
This is consistent with the fact that we
expect Witten's effect \cite{witten} to take place for magnetic
monopoles - a monopole with a magnetic charge $h$ 
becomes a dyon with  the electric charge
\begin{eqnarray}
\label{electric}
e= \Big ( {\theta g_{\rm YM}^2\over 16\pi^2} \Big )~h~
\end{eqnarray}
in the presence of the $\theta$ angle \cite {witten,tHooft}. 
Thus, the fact that the interaction force 
derived from the above matrix quantum mechanics depends
non-trivially on the $\theta$-angle gives additional 
evidence that 't Hooft's QCD monopoles might indeed be described 
by the former. Dyons in this case can (very roughly) be thought of 
as complicated bound states of a monopole and off-diagonal gluons.  

{}Let us also point out that the string tension
$T_s$ is invariant under the S-duality transformation
$\tau\rightarrow1/\tau$. On the other hand, at first it might appear
strange that it is not invariant under the shift
$\theta\rightarrow\theta+2\pi$. This is, however, expected as the electric
charge $e$ given by (\ref{electric}) is not invariant under such shifts
either;  the charge (\ref {electric})
gets shifted by a fundamental unit of the ``electric charge''
(which in our notations is $g^2_{\rm YM} h/ 8\pi$), {\em i.e.}, 
$e\rightarrow e+g^2_{\rm YM} h/8 \pi$. 
This can be interpreted as follows: 
't Hooft's dyons  at $\theta$ can be viewed as a bound
state of the corresponding dyon  at $\theta-2\pi$ and a gluon 
\cite{tHooft}.  

{}The work of G.G. was  supported by the grant
NSF PHY-94-23002. The work of Z.K. was supported in part by the 
grant NSF PHY-96-02074, and the DOE 1994 OJI award. Z.K. would also 
like to thank Albert and Ribena Yu for financial support.

\end{document}